\documentclass[numberedappendix]{emulateapj}
\usepackage[normalem]{ulem}

\def\msun{\rm M_{\sun}}

\def\micron{$\mu$m}

\def\mdot{\rm \dot{M}}

\begin{document}


\bibliographystyle{apj}

\shorttitle{GM Aur Protoplanetary Disk}

\shortauthors{Hughes et al.}

\slugcomment{Accepted for publication in ApJ: March 25, 2009}
\title{
A Spatially Resolved Inner Hole in the Disk around GM Aurigae
}

\author{A. Meredith Hughes\altaffilmark{1}, 
Sean M. Andrews\altaffilmark{1,2}, 
Catherine Espaillat\altaffilmark{3}, 
David J. Wilner\altaffilmark{1}, 
Nuria Calvet\altaffilmark{3}, \\ 
Paola D'Alessio\altaffilmark{4}, 
Chunhua Qi\altaffilmark{1}, 
Jonathan P. Williams\altaffilmark{5},
and Michiel R. Hogerheijde\altaffilmark{6}}
\altaffiltext{1}{Harvard-Smithsonian Center for Astrophysics, 60 Garden Street, Cambridge, MA 02138; mhughes, sandrews, dwilner, cqi$@$cfa.harvard.edu}
\altaffiltext{2}{Hubble Fellow}
\altaffiltext{3}{Department of Astronomy, University of Michigan, 830 Dennison Building, 500 Church Street, Ann Arbor, MI 48109; ccespa, ncalvet$@$umich.edu}
\altaffiltext{4}{Centro de Radioastronom{\'{\i}}a y Astrof{\'{\i}}sica, Universidad Nacional Aut{\'{o}}noma de M{\'{e}}xico, 58089 Morelia, Michoac{\'{a}}n, Mexico; p.dalessio@astrosmo.unam.mx}
\altaffiltext{5}{University of Hawaii Institute for Astronomy, 2680 Woodlawn Drive, Honolulu, HI 96822; jpw$@$ifa.hawaii.edu}
\altaffiltext{6}{Leiden Observatory, Leiden University, P.O. Box 9513, 2300 RA Leiden, Netherlands; michiel$@$strw.leidenuniv.nl}

\begin{abstract}
We present 0\farcs3 resolution observations of the disk around GM Aurigae 
with the Submillimeter Array (SMA) at a wavelength of 860\,$\mu$m and with the 
Plateau de Bure Interferometer at a wavelength of 1.3\,mm.  These observations
probe the distribution of disk material on spatial scales commensurate 
with the size of the inner hole predicted by models of the spectral energy 
distribution.  The data clearly indicate a sharp decrease in millimeter 
optical depth at the disk center, consistent with a deficit of material at 
distances less than $\sim$20\,AU from the star.  We refine the accretion disk
model of \citet{cal05} based on the unresolved spectral energy distribution
(SED) and demonstrate that it reproduces well the spatially resolved 
millimeter continuum data at both available wavelengths.  We also present 
complementary SMA observations of CO $J$=3$-$2 and $J$=2$-$1 emission from 
the disk at 2\arcsec\ resolution.  The observed CO morphology is consistent 
with the continuum model prediction, with two significant deviations: (1) the 
emission displays a larger CO $J$=3$-$2/$J$=2$-$1 line ratio than predicted, 
which may indicate additional heating of gas in the upper disk layers; and 
(2) the position angle of the kinematic rotation pattern differs by 
$11^\circ \pm 2^\circ$ from that measured at smaller scales from the dust 
continuum, which may indicate the presence of a warp.  We note that 
photoevaporation, grain growth, and binarity are unlikely mechanisms for 
inducing the observed sharp decrease in opacity or surface density at 
the disk center.  The inner hole plausibly results from the dynamical 
influence of a planet on the disk material.  Warping induced by a planet 
could also potentially explain the difference in position angle between the 
continuum and CO data sets.
\end{abstract}


\section{Introduction}

Understanding of the planet formation process is intimately tied to 
knowledge of the structure and evolution of protoplanetary disks.  
Of particular importance is how and when in the lifetime of the disk 
its constituent material is cleared, which provides clues to how and 
when planets may be assembled.  While observations suggest that the inner 
and outer dust disk disperse nearly simultaneously \citep[e.g.][]{skr90,wol96,
and05}, it is not clear which physical mechanism(s) drives this process, or 
the details of how it progresses.  Possible dispersal mechanisms, of which 
several may come into play over the lifetime of a disk, include a 
drop in dust opacity due to grain growth \citep[e.g.] []{str89,dul05}, 
photoevaporation of material by energetic stellar radiation 
\citep[e.g.][]{cla01}, photophoretic effects of gas on dust grains 
\citep{kra05}, inside-out evacuation via the magnetorotational 
instability \citep{chi07}, and the dynamical interaction of giant planets 
with natal disk material \citep[e.g.][]{lin86,bry99}.  Observing the 
distribution of gas and dust in disks allows us to evaluate the roles of 
these disk clearing mechanisms.

One particular class of systems, those with ``transitional'' disks 
\citep[e.g.][]{str89,skr90}, have become central to our understanding 
of disk clearing.  These disks exhibit a spectral energy distribution 
(SED) morphology with a deficit in the near- to mid-infrared excess over the 
photosphere consistent with a depletion of warm dust near the star.  The advent 
of the {\it Spitzer} Space Telescope has allowed detailed measurement of 
mid-infrared spectra with unprecedented quality and quantity.  Combined 
with simultaneous advances in disk modeling that can now reproduce in 
detail the SED features \citep[e.g.][]{dal99,dal01,dul02,dal06}, these 
observations have revolutionized the study of disk structure.  
However, such studies 
rely entirely on SED deficits whose interpretations are not unique, since 
effects of geometry and opacity can mimic the signature of disk clearing 
\citep{bos96,chi99}.  

Spatially resolved observations are crucial for confirming the structures 
inferred from disk SEDs.  High resolution imaging at millimeter wavelengths 
is especially important because dust opacities are low, and the disk mass 
distribution can be determined in a straightforward way for an assumed 
opacity.  Millimeter observations also avoid many of the complications 
present at shorter wavelengths, including large optical depths, spectral 
features, and contrast with the central star.  Several recent millimeter 
studies have resolved inner emission cavities for disks with infrared SED 
deficits through direct imaging observations, e.g.  TW Hya \citep{cal02,hug07}, 
LkH$\alpha$~330 \citep{bro07,bro08}, and LkCa 15 \citep{pie07,esp08}.  
These observations unambiguously associate infrared SED deficits with a 
sharp drop in millimeter optical depth in the disk center.  More information 
is needed to determine whether the low optical depth is a result of 
decreased surface density or opacity.

GM Aurigae is a prototypical example of a star host to a ``transitional''
disk.  The $\sim$1-5 Myr old T Tauri star \citep{sim95,gul98} of spectral 
type K5 is located at a distance of 140\,pc in the 
Taurus-Auriga molecular complex \citep{ber06}, and its brightness and 
relative isolation from intervening cloud material have enabled a suite of 
observational studies of its disk properties.  The presence of circumstellar 
dust emitting at millimeter wavelengths was first inferred by \citet{wei89}, 
and the disk structure was subsequently resolved in the $^{13}$CO $J$=2--1 
transition by \citet{koe93}.  Their arcsecond-resolution mapping of the 
gas disk revealed gaseous material in rotation about the central star.
Assuming a Keplerian rotation pattern allowed a determination 
of the dynamical mass for the central star of 0.8 M$_\sun$.  Further modeling 
of the structure and dynamics of the disk was carried out by \citet{dut98}, 
using higher-resolution $^{12}$CO $J$=2--1 observations.  Scattered light 
images revealed a dust disk inclined by 50-56\degr\ extending to radii 
$\sim300$\,AU from the star \citep{sta97,sch03}.

Efforts to model the SED of GM Aurigae have long indicated the presence of 
an inner hole, and estimates of its size have grown over the years as the 
quality of data and models have improved.  In the early 1990s, the low 
12\,$\mu$m flux led to $\sim0.5$\,AU estimates of the inner disk radius 
\citep{mar92,koe93}.  That value was later increased to 4.8\,AU by 
\citet{chi99} in the context of hydrostatic radiative equilibrium models, 
and a putative planet at a distance of 2.5\,AU from the star was shown to
be capable of clearing an inner hole of this extent using simulations of the
relevant hydrodynamics \citep{ric03}.  
With the aid of a ground-based mid-IR spectrum, \citet{ber04} increased 
the gap size estimate to 6.5\,AU, and subsequently \citet{cal05} inferred 
an inner hole radius of 24\,AU using a {\it Spitzer} IRS spectrum in 
combination with sophisticated disk structure models.  Recently, \citet{dut08} 
have argued for a 19$\pm$4\,AU inner hole in the gas distribution, using 
combined observations of several different molecular line tracers.  Like the 
SED-based measurements, their method is indirect: they use a model of the 
disk in Keplerian rotation to associate a lack of high-velocity molecular 
gas with a deficit of material in the inner disk.

We present interferometric observations at 860\,$\mu$m from the Submillimeter 
Array\footnote{The Submillimeter Array is a joint project between the 
Smithsonian Astrophysical Observatory and the Academica Sinica Institute of 
Astronomy and Astrophysics and is funded by the Smithsonian Institution and the
Academica Sinica.} and 1.3\,mm from the Plateau de Bure 
Interferometer\footnote{Based on observations carried out with the 
IRAM Plateau de Bure Interferometer. IRAM is supported by INSU/CNRS (France), 
MPG (Germany) and IGN (Spain).} that probe disk material on scales 
commensurate with the 24\,AU inner disk radius inferred from the SED.  These
data allow us to directly resolve the inner hole in the GM Aur disk for the 
first time.  We describe the observations in \S\ref{sec:obs} and present the 
dual-wavelength continuum data in \S\ref{sec:continuum}.  We also present 
observations of the molecular gas disk in the CO $J$=3--2 and $J$=2--1 lines 
that allow us to study disk kinematics in \S\ref{sec:CO}.  We use these data 
to investigate disk structure in the context of the SED-based models of 
\citet{cal05}, described in \S\ref{sec:model}.  Implications for the disk 
structure and evolutionary status are discussed in \S\ref{sec:discussion}.

\section{Observations and Data Reduction}
\label{sec:obs}

\begin{figure*}[ht]
\epsscale{1.0}
\plotone{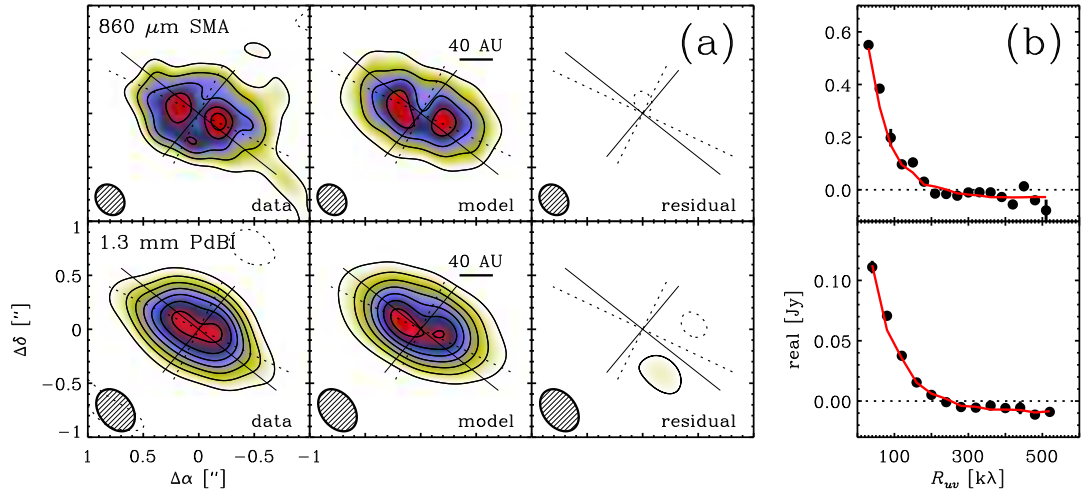}
\figcaption{Continuum emission from the disk around GM Aur at wavelengths of 
860\,$\mu$m observed with the SMA ({\it top}) and 1.3\,mm observed with PdBI 
({\it bottom}).  The data are displayed in both the image ({\it a}) and 
Fourier ({\it b}) domains.  In the image domain ({\it a}), the observed 
brightness distribution at each wavelength ({\it left}) is compared with the 
model prediction ({\it center}; see \S\ref{sec:modelcont} for model details), 
and the residuals are also shown ({\it right}).  In the data and model frames, 
the contours are $[3, 6, 9, ... ] \times $ the rms noise (3.5\,mJy\,beam$^{-1}$ 
at 860\,$\mu$m and 0.75\,mJy\,beam$^{-1}$ at 1.3\,mm).  In the residual frame, 
the contours start at 2$\sigma$ and are never greater than 3$\sigma$.   The 
synthesized beam sizes and orientations for the two maps are, respectively, 
$0\farcs30\times0\farcs24$ at a position angle of 34\degr\ and 
$0\farcs43\times0\farcs30$ at a position angle of 35\degr.  Two sets of axes
are shown: the dotted line indicates the position angle of the double-peaked
continuum emission, while the solid line indicates the best-fit position angle
of the CO emission (see \S\ref{sec:CO} for details).  In the Fourier 
domain ({\it b}), the visibilities are averaged in bins of deprojected 
{\it u-v} distance from the disk center, and compared with the model 
prediction (red line).  The inner hole in the GM Aur disk is clearly observed 
at both wavelengths, as a double-peaked emission structure in the image 
domain or as a null in the visibility function in the Fourier domain. 
\label{fig:continuum}}
\end{figure*}

The GM Aur disk was observed with the 8-element (each with a 6\,m diameter) 
Submillimeter Array \citep[SMA;][]{ho04} in the very extended (68-509\,m 
baselines) and compact (16-70\,m baselines) configurations on 2005 November 5 
and 26, respectively.  Observing conditions on both nights were excellent, with 
$\sim$1\,mm of precipitable water vapor and good phase stability.  Double 
sideband receivers were tuned to a central frequency of 349.935\,GHz 
(857\,$\mu$m), with each 2\,GHz-wide sideband centered $\pm$5\,GHz from that 
value.  The SMA correlator was configured to observe the CO $J$=3$-$2 
(345.796\,GHz) and HCN $J$=4$-$3 (354.505\,GHz) transitions with a velocity 
resolution of 0.18\,km s$^{-1}$.  No HCN was detected, with a 3$\sigma$ upper 
limit of 0.9\,Jy\,beam$^{-1}$ in the 2\farcs2$\times$1\farcs9 synthesized beam. 
The observing sequence alternated between 
GM Aur and the two gain calibrators 3C 84 and 3C 111. The data were edited and 
calibrated using the MIR software package.\footnote{See \url{http://cfa-www.harvard.edu/$\sim$cqi/mircook.html}.}  The passband response was calibrated using 
observations of Saturn (compact configuration) or the bright quasars 3C 273 and 
3C 454.3 (very extended configuration).  The amplitude scale was determined 
by bootstrapping observations of Uranus and these bright quasars, and is 
expected to be accurate at the $\sim$10\%\ level.  Antenna-based gain 
calibration was conducted using 3C 111, while the 3C 84 observations were 
used to check on the quality of the phase transfer.  We infer that the 
``seeing" induced on the very extended observations by phase noise and small 
baseline errors is small, $\lesssim$0\farcs1.  Wideband continuum channels 
from both sidebands and configurations were combined.  The derived 870\,$\mu$m
flux of GM Aur is $640\pm60$\,mJy.

Additional SMA observations in the extended (28-226\,m) and sub-compact 
(6-69\,m baselines) configurations were conducted on 2006 December 10 and 
2007 September 14, respectively, with a central frequency of 224.702\,GHz 
(1335\,$\mu$m).  While the sub-compact observations were conducted in typical 
weather conditions for this band (2.5\,mm of water vapor), the extended data 
were obtained in better conditions similar to those for the higher frequency 
observations described above.  The correlator was configured to simultaneously 
cover the $J$=2$-$1 transitions of CO (230.538\,GHz), $^{13}$CO (220.399\,GHz), 
and C$^{18}$O (219.560\,GHz) with a velocity resolution of 
$\sim$0.28\,km s$^{-1}$.  The calibrations were performed as above. 

GM Aurigae was also observed with the 6-element (each with a 15\,m diameter) 
Plateau de Bure Interferometer (PdBI) in the A configuration (up to 750\,m 
baselines) on 2006 January 15.  Observing conditions were excellent, with 
atmospheric phase noise generating a seeing disk of $\lesssim$0.2\arcsec.  The 
PdBI dual-receiver system was set to observe the 110.201\,GHz (2.7\,mm) and 
230.538\,GHz (1.3\,mm) continuum simultaneously.  As with the SMA data, 
observations alternated between GM Aur and two gain calibrators, 3C 111 and 
J0528+134.  The data were edited and calibrated using the GILDAS package 
\citep{pet05}.  The passband responses and amplitude scales were calibrated 
with observations of 3C 454.3 and MWC 349, respectively.  The derived 1.3 and
2.7\,mm fluxes of GM Aur are $180\pm20$ and $21\pm2$\,mJy.

The standard tasks of Fourier inverting the visibilities, deconvolution with 
the CLEAN algorithm, and restoration with a synthesized beam were conducted 
with the MIRIAD software package.  A high spatial resolution image of the 
860\,$\mu$m continuum emission from the SMA data was created with a Briggs 
robust = 1.0 weighting scheme for the visibilities, excluding projected 
baselines $\le 70$\,k$\lambda$, resulting in a synthesized beam FWHM of
$0\farcs30\times0\farcs24$ at a position angle of 34\degr.  A similar 
image of the 1.3\,mm continuum emission with a synthesized beam FWHM of 
$0\farcs43\times0\farcs30$ at a position angle of 35\degr\ was generated 
from the PdBI data using natural weighting (robust = 2.0).  Table~\ref{tab:obs} 
summarizes the line and continuum observational parameters.

\begin{table*}
\caption{Observational parameters for GM Aur}
\begin{tabular}{lccccc}
\hline
 & & & \multicolumn{3}{c}{Continuum} \\
\cline{4-6}
Parameter & $^{12}$CO $J$=3--2 & $^{12}$CO $J$=2--1 & 860\,$\mu$m & 1.3\,mm & 2.7\,mm \\
\hline
\hline
Rest Frequency (GHz) & 345.796 & 230.538 & 349.935 & 230.538 & 110.201 \\
Channel Width & 0.18\,km\,s$^{-1}$ & 0.28\,km\,s$^{-1}$ & $2 \times 2$\,GHz & $2\times548$\,MHz & 548\,MHz \\
Beam Size (FWHM) & 2\farcs2$\times$1\farcs9 & 2\farcs1$\times$1\farcs4 & 0\farcs30$\times$0\farcs24 & 0\farcs43$\times$0\farcs30 & 0\farcs93$\times$0\farcs60 \\
~~~PA & 14\degr & 56\degr & 34\degr & 35\degr & 31\degr \\
RMS noise (mJy\,beam$^{-1}$) & 310 & 90 & 3.5 & 0.75 & 0.25 \\
Peak Flux Density (mJy\,beam$^{-1}$) & $6700\pm300$ & $2400\pm100$ & $59\pm4$ & $16.6\pm0.8$ & $10.3\pm0.3$ \\
Integrated Continuum Flux (mJy) & -- & -- & $640\pm60$ & $180\pm20$ & $21\pm2$ \\
Integrated Line Intensity (Jy\,km\,s$^{-1}$) & 29 & 37 & -- & -- & -- \\
\hline
\end{tabular}
\label{tab:obs}
\end{table*}

\section{Results}

\subsection{Millimeter Continuum Emission}
\label{sec:continuum}

Figure \ref{fig:continuum} shows the results of the SMA and PdBI continuum
observations in both the image and Fourier domains.  The presence of an 
inner hole in the GM Aur disk, as predicted by models of the SED, is clearly 
indicated both by the double-peaked emission structure in the image and by 
the null in the visibility data.  The double-peaked emission structure points 
to a deficit of flux near the disk center; the null in the visibility 
function, or the location at which the real part of the visibilities change 
sign, similarly reflects a decrease in flux at small angular scales.  
The resolution of the 2.7\,mm data from the PdBI was insufficient to provide 
about the inner hole.

The maps in the left panel of Fig.~\ref{fig:continuum} show a double-peaked 
brightness distribution at both wavelengths, with peak flux densities of 
$59 \pm 4$ mJy beam$^{-1}$ at 860\,$\mu$m and $16.6 \pm 0.3$ mJy beam$^{-1}$ 
at 1.3\,mm.  For all but the most edge-on viewing geometries 
\citep[e.g.][]{wol08}, a continuous density distribution extending in to the 
dust destruction radius \citep[$\sim$0.05-0.1\,AU; ][]{ise06} would be expected to result in a 
centrally-peaked brightness distribution.  In the case of GM Aurigae, the 
double-peaked emission structure is a geometric effect due to the truncation 
of disk material at a much larger radius, viewed at an intermediate 
inclination of 50-56$^\circ$ \citep{dut98,dut08}: the region of highest 
density is near the inner disk edge, with a large column density of optically 
thin material in this ring effectively generating limb brightening at the 
inner edge of the outer disk, at two points along the disk major axis.  

The size of the inner hole can be roughly estimated by the separation of the 
emission peaks, although the peak separation will also depend on the 
brightness of the directly-illuminated inner edge of the outer disk relative 
to the extended disk component \citep{hug07}.  The separation of the peaks 
in the 860 $\mu$m image is $0\farcs38 \pm 0\farcs03$, corresponding to a 
physical diameter of $53 \pm 4$ AU (radius $27 \pm 2$ AU) at a distance of 
140\,pc.  A position angle of 66$^\circ$ is estimated by the orientation of
a line that bisects the two peaks, although a more robust value of $64^\circ 
\pm 2^\circ$ is derived in \S\ref{sec:modelcont} below.  Since the peaks are 
not distinctly separated in the 1.3\,mm image, the same estimate cannot be 
made, but the position angle is clearly consistent with that derived from the 
860\,$\mu$m visibilities and indicated by the perpendicular dashed lines in 
Fig.~\ref{fig:continuum}.

The presence of an inner hole is also evident from the visibilities 
displayed in the right panel of Fig.~\ref{fig:continuum}.  The real part of 
the complex visibilities have been averaged in concentric annuli of 
deprojected $(u,v)$ distance from the disk center.  For details of the 
deprojection process, see \citet{lay97}.  As discussed in the appendix
of \citet{hug07}, the presence of a null in the visibility function indicates
a sharp decrease in flux at a radius corresponding roughly to the angular 
scale of the null position.  The precise position of the null depends 
primarily on the angular size of the inner hole, but also on the radial 
gradients of the surface density and temperature distribution and the 
relative brightness of the directly illuminated wall at the inner edge of 
the outer disk.  In a standard power-law parameterization, the disk temperature
$T$ and surface density $\Sigma$ vary inversely with radius as $\Sigma \propto 
R^{-p}$ and $T \propto R^{-q}$.  Neglecting the emission from the wall
and assuming standard values of $p=1.0$ and $q=0.5$, 
expected for a typical viscous disk with constant $\alpha$ \citep{har98} and 
consistent with previous studies of the GM Aur disk \citep{dut98,and07,hug08}, 
we may obtain a rough estimate of the size of the inner hole using the observed 
null position and Eq.~A9 from \citet{hug07}: $\mathcal{R}_{\mathrm{null}} 
(\textrm{k}\lambda) = (1 \textrm{ AU} / R_{\mathrm{hole}}) (D_{\mathrm{source}} 
/ 100 \textrm{ pc}) [2618 + 1059(p + q)]$.  A polynomial curve fit to 
the visibilities yields a null position of 190 k$\lambda$ at 860\,$\mu$m
and 224 k$\lambda$ at 1.3\,mm, which correspond to inner hole radii of 31 
and 26 AU, respectively.  However, these estimates are uncertain to within
$\sim$30\%, as the data are consistent with a broad range of null positions.  
We therefore turn to a more sophisticated modeling procedure described in 
\S\ref{sec:modelcont} below.

\subsection{CO Channel and Moment Maps}
\label{sec:CO}

\begin{figure*}[t]
\epsscale{1.0}
\plotone{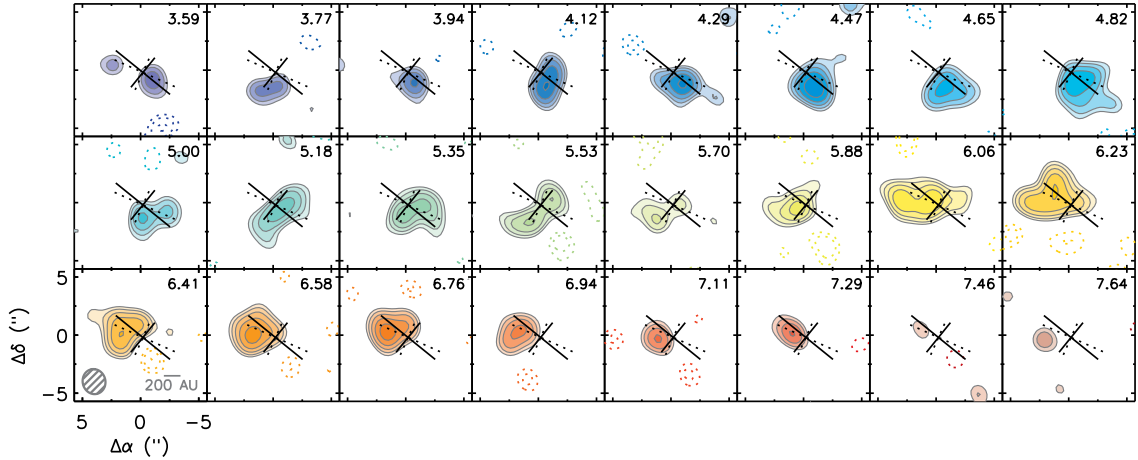}
\figcaption{Channel maps of CO $J$=3$-$2 emission from the GM Aur disk.  Contour
levels start at 0.61 Jy (2 times the rms noise) and increase by factors of
$\sqrt{2}$.  LSR velocity is indicated by color and quoted in the upper right
of each panel.  The synthesized beam (2$\farcs$2$\times$1$\farcs$9 at a PA of
14$^\circ$) and physical scale are indicated in the lower left panel.  Two 
sets of axes are shown: the dotted line indicates the position angle of the 
double-peaked continuum emission, while the solid line indicates the best-fit 
position angle of the CO emission.  
\label{fig:co32}
}
\end{figure*}
\begin{figure*}[t]
\epsscale{1.0}
\plotone{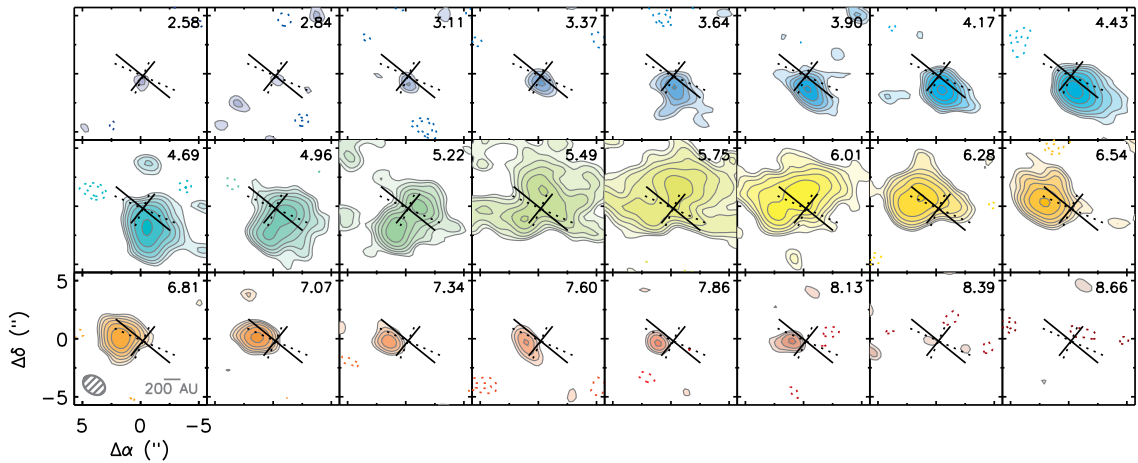}
\figcaption{Channel maps of CO $J$=2$-$1 emission from the GM Aur disk.  Contour 
levels start at 0.17 Jy (2 times the rms noise) and increase by factors of 
$\sqrt{2}$.  LSR velocity is indicated by color and quoted in the upper right 
of each panel.  The synthesized beam (2$\farcs$1$\times$1$\farcs$4 at a PA of 
56$^\circ$) and physical scale are indicated in the lower left panel.  Two 
sets of axes are shown: the dotted line indicates the position angle of the 
double-peaked continuum emission, while the solid line indicates the best-fit 
position angle of the CO emission.  Cloud contamination is evident in at least 
the central four channels.  
\label{fig:co21}
}
\end{figure*}

\begin{figure}[t]
\epsscale{1.0}
\plotone{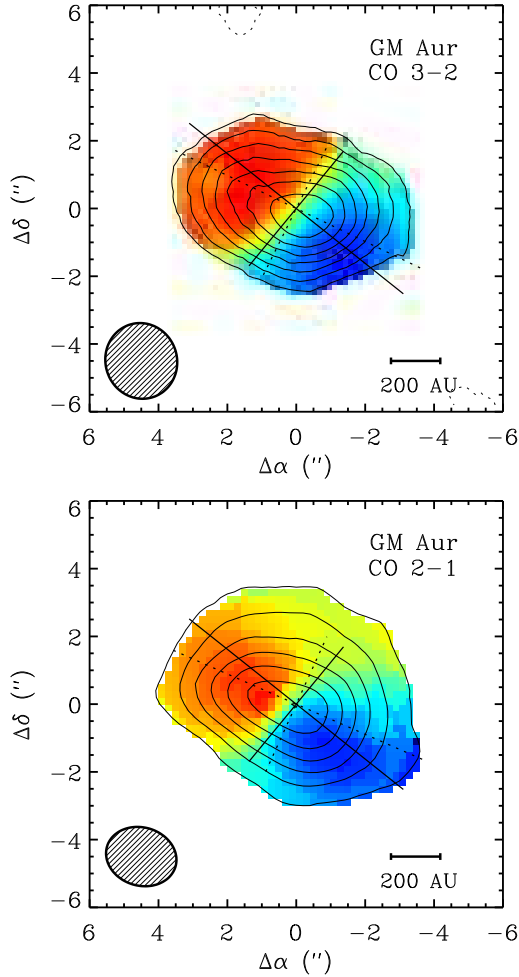}
\figcaption{Zeroeth ({\it contours}) and first ({colors}) moment map of the
CO $J$=3$-$2 ({\it top}) and $J$=2$-$1 ({\it bottom}) data in Figs.~\ref{fig:co32} and
\ref{fig:co21}.  The dotted line indicates the position angle of the 
double-peaked continuum emission, while the solid line indicates the 
best-fit position angle of the CO emission.  The zeroeth moment contours
are well aligned with the latter, while the isovelocity contours of the first 
moment map are more consistent with the former.  Cloud contamination is 
evident in the CO $J$=2$-$1 map in the northwest region along the disk minor axis.
\label{fig:momfig}
}
\end{figure}

Figures \ref{fig:co32}--\ref{fig:momfig} display the new SMA observations
of CO emission from the GM Aur disk.  Figures \ref{fig:co32} and 
\ref{fig:co21} show channel maps with contours starting at twice the 
rms noise level and increasing by factors of $\sqrt{2}$, while Figure 
\ref{fig:momfig} displays the zeroeth (contours) and first (color) 
moments of the data: these are the velocity-integrated intensity and 
intensity-weighted velocities, respectively.  The peak flux density is 
$6.7\pm0.3$\,Jy\,beam$^{-1}$ in the CO $J$=3--2 line and 
$2.4\pm0.1$\,Jy\,beam$^{-1}$ in the CO $J$=2--1 line, with integrated fluxes 
of 9.4\,Jy\,km\,s$^{-1}$ and 21.2\,Jy\,km\,s$^{-1}$, respectively (although 
emission from extended ambient cloud material is likely to increase the 
CO $J$=2--1 integrated flux over that originating from the disk alone).  
The channel and moment maps are broadly consistent with the expected 
kinematic pattern for material in Keplerian rotation about the central star, 
substantially inclined to our line of sight \citep[as in][]{dut98,sim00}.  

The short-baseline spatial frequencies in the $(u,v)$ plane provided by the 
subcompact configuration of the SMA during our observations of the $J$=2$-$1 
transition are sensitive to emission on the largest spatial scales.  These
short antenna spacings reveal the severity of the cloud contamination to an 
extent not possible with previous data.  The contamination is evident as an 
extended halo around the disk emission in the central channels of the $J$=2$-$1 
channel maps near LSR velocities of 5-6\,km\,s$^{-1}$ (Fig.~\ref{fig:co21}).
It is also evident in the moment map (Fig.~\ref{fig:momfig}) as an elongation 
of emission near the systemic velocity (green-yellow) to the northwest along 
the disk minor axis.  This contamination indicates that caution
must be exercised when deriving kinematic information from the CO lines, 
particularly the central channels.  Spatial filtering by the interferometer
does not ameliorate cloud contamination in an abundant, easily-excited, 
high-optical depth tracer like CO$J$=2--1.  The $J$=3$-$2 line appears less 
contaminated than $J$=2$-$1 (Figs.~\ref{fig:co32} and \ref{fig:momfig}), 
although similarly short antenna spacings (8-43\,m) are not present 
in this data set.  Nevertheless, we expect less cloud contamination in the
$J$=3$-$2 transition, since the temperature of the cloud will be lower than 
that of the disk and will therefore populate the upper rotational levels of 
the CO molecule less efficiently.  The cloud contamination prevents 
detection of self-absorption in the central channels of the CO $J$=2--1 
channel maps along the near (northwest) edge of the disk \citep[as 
determined by scattered light observations; see ][]{sch03}.  \citet{dut98} 
report self-absorption along the southeast edge, but our observations 
suggest that this brightness asymmetry may be due to cloud contamination.  
It is also possible that the contamination is due to a residual envelope,
although we are unable to determine the large-scale structure of the extended
line emission with our interferometric data.

In all figures, the disk orientation based on the position angle of 64\degr\
derived from the continuum emission (Fig.~\ref{fig:continuum} and 
\S\ref{sec:modelcont}) is plotted over the CO emission as a set of crossed 
dashed lines, with the relative extent of the major and minor axes (based on 
the inclination angle of 55\degr) indicated by the length of the perpendicular 
lines.  The position angle of 51\degr\ derived by \citet{dut98} from fitting 
the CO $J$=2$-$1 emission, consistent with our own $J$=3$-$2 and $J$=2$-$1 
observations, is illustrated by the solid line.  Note that the position angle 
of the CO emission differs slightly from the position angle of the continuum 
emission, by $11^\circ \pm 2^\circ$ (see \S\ref{sec:modelcont}).  The trend 
is clear for both transitions, but more obvious in the less-contaminated 
$J$=3$-$2 transition.  Note that the position angle for the CO emission is 
derived entirely from the rotation pattern (evident in the isovelocity 
contours) and not from the geometry of the integrated CO emission: the 
integrated emission appears to match the position angle from the continuum 
emission reasonably well.  We do not observe the isophote twisting in 
integrated CO emission seen by \citet{dut98}.  The cloud contamination and 
differences in antenna spacings may play a role.

\section{Disk Structure Models}
\label{sec:model}

\subsection{Updated SED Model}
\label{sec:modelcont}

Here we revisit the broadband SED modeling of GM Aur presented by
\citet{cal05}. Taking into consideration new observational constraints 
at sub-millimeter and millimeter wavelengths, we use the irradiated 
accretion disk models of \citet{dal05,dal06} to re-derive 
the properties of the outer disk of GM Aur and its inner, truncated 
edge or ``wall.'' Our grain-size distribution follows a power-law of 
$a^{-3.5}$, where $a$ is the grain radius.  We assume ISM-sized 
grains in the upper layers of the disk and accordingly 
adopt $a_{min}$=0.005\,{\micron} and $a_{max}$=0.25\,{\micron} \citep{dra84}.  
Closer to the disk midplane grains have a maximum size of 1\,mm. Input 
parameters for the outer disk include the stellar properties, the mass 
accretion rate, the viscosity parameter ($\alpha$), and the settling 
parameter ($\epsilon$) which measures the dust-to-gas mass ratio in the 
upper layers of the disk relative to the standard dust-to-gas mass ratio. 
Following \citet{cal05}, we adopt the same extinction, distance, inclination, 
dust grain opacities, and stellar properties (i.e. luminosity, radius, and 
temperature; see Table \ref{tab:prop}).  We use a mass accretion rate of 
7.2$\times$10$^{-9}$\,$\msun$\,yr$^{-1}$ which was derived using 
{\it HST} STIS spectra by \citet{ing09}, in contrast to the
value of $10^{-8}$\,M$_\sun$\,yr$^{-1}$ derived from veiling measurements in 
\citet{cal05}.  We assume an outer disk radius of 300\,AU, which matches the
observed extent of scattered light from the dust disk \citep{sch03} and 
previous fits to the continuum emission \citep{hug08}, as well as the 
short-baseline data presented here. 

\begin{figure*}[ht]
\epsscale{1.0}
\plottwo{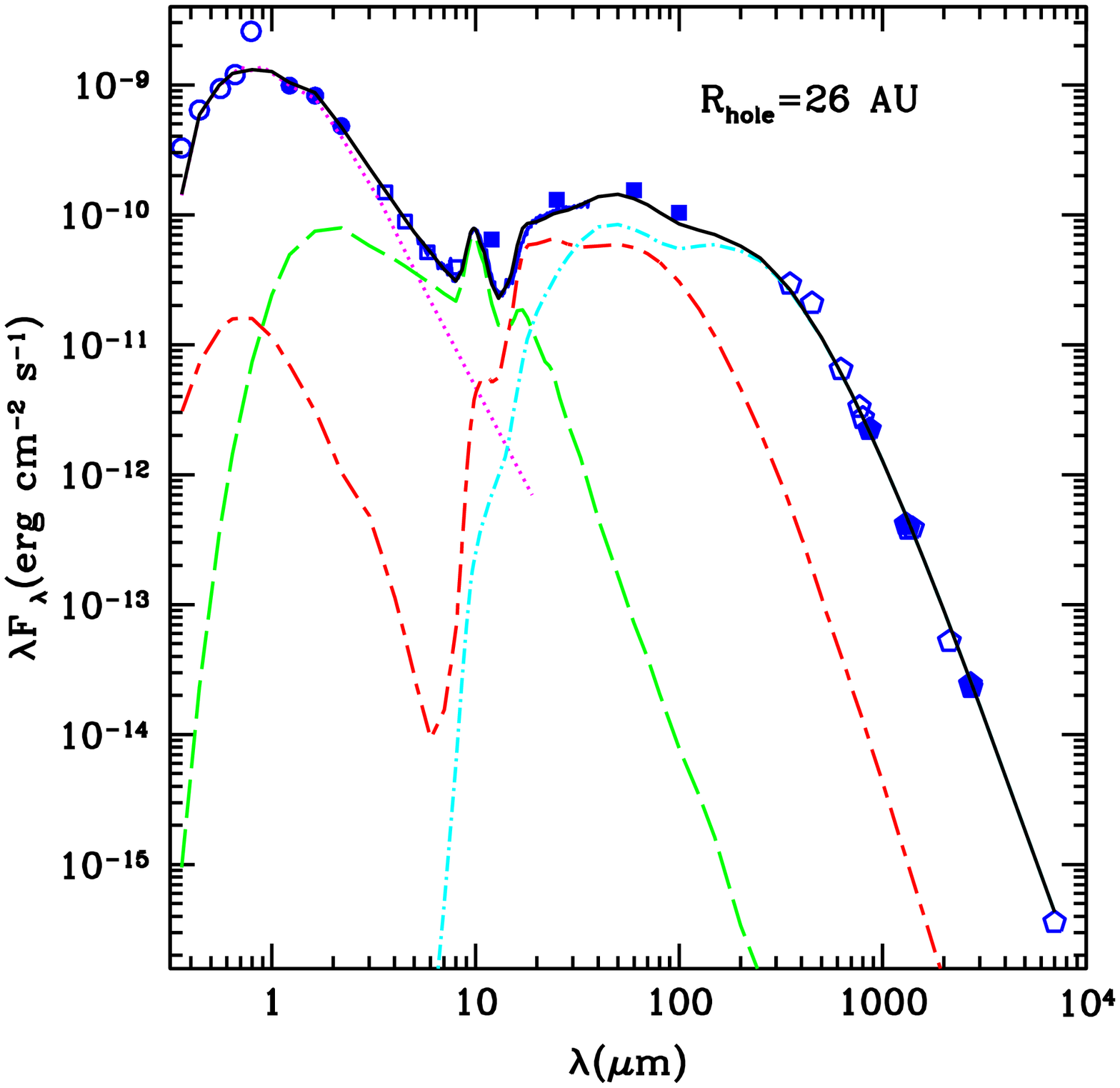}{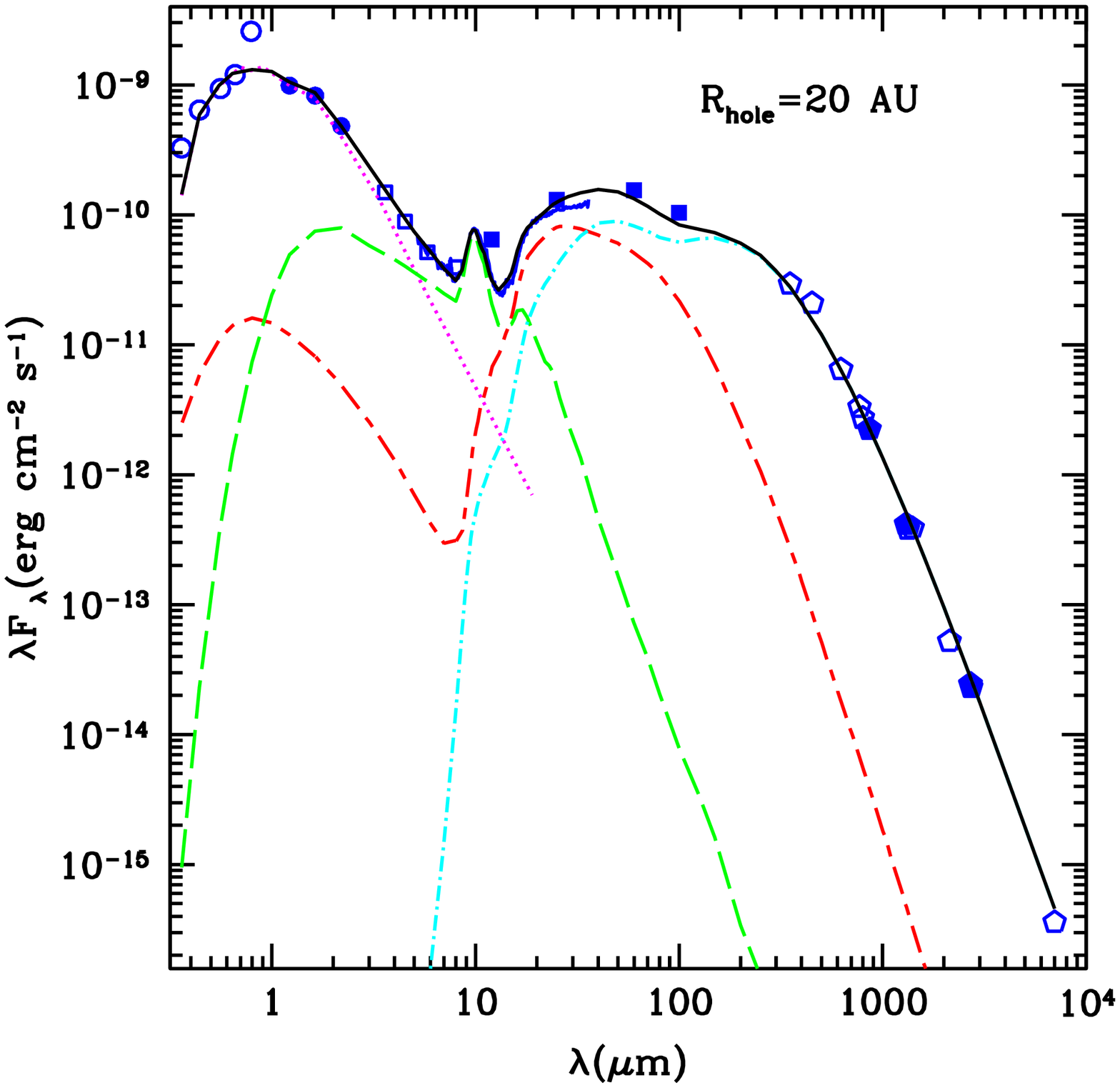}
\figcaption{Model of the SED of GM Aur using the method of 
\citet{dal05,dal06}.  The final model of the SED alone has an inner disk 
hole of 26 AU ({\it left}), while the model that best reproduces the resolved
millimeter-wavelength visibilities has a hole of radius 20\,AU ({\it right}).  
See \S\ref{sec:model} for model details.  We show optical \citep[open 
circles;][]{ken95}, 2MASS (closed circles), IRAC \citep[open squares;][]{
har05}, and IRAS \citep[closed squares;][]{wea92} data and a {\it Spitzer} 
IRS spectrum \citep{cal05}.  Open pentagons represent millimeter observations 
obtained from \citet{and05,bec91,dut98,kit02,koe93,loo00,rod06,wei89}.  Closed 
pentagons are from this work.  The final model (solid line) includes 
the following components: stellar photosphere (dotted line), optically
thin dust region (long-dashed line), disk wall (short-long dashed line), 
outer disk (dot-dashed line).  The peak at $\sim$1{\micron} from the wall 
emission is due to scattered light.  While the 20\,AU model does not fit 
the IRS spectrum as well between $\sim$20--40 {\micron} as the 26\,AU model, 
it reproduces the millimeter continuum emission very well at both 860\,$\mu$m 
and 1.3\,mm (Fig.~\ref{fig:continuum}).
\label{fig:catherine}}
\end{figure*}

In order to reproduce the outer disk component of the SED, we vary $\epsilon$ 
and $\alpha$ (Figure \ref{fig:catherine}).  As described in \citep{cal05}, 
$\alpha$ effectively determines the mass surface density distribution and 
therefore the disk mass, which is best reflected by the longest-wavelength 
SED points.  The value of $\epsilon$ has the greatest effect on the slope 
of the SED beyond 100\,$\mu$m.  With the new millimeter data we find 
$\epsilon$=0.5, indicating less settling than reported previously.  We also 
find $\alpha$=0.002 and a more massive outer disk of 0.16\,$\msun$.  This 
mass is significantly larger than an estimate based on the 860\,$\mu$m and 
1.3\,mm flux measurements using opacities from \citet{bec90}, which yields 
$\sim$0.04\,M$_\sun$, and is only marginally Toomre stable at 300\,AU 
(Q$\sim$1.1).   The outer disk model uses an opacity of $\sim$0.1\,cm$^{2}$ 
g$^{-1}$ at 1\,mm \citep{dal01} which is about four times lower than that 
derived from the \citet{bec90} opacities, accounting for the discrepancy in 
mass.  Within the inner disk hole, there are 1.1$\times$10$^{-11}$\,$\msun$ 
of optically thin small dust grains, which account for the 10 {\micron} 
emission and the near-IR excess. The mass in solids could be much larger than
this mass if pebbles, rocks, or even planetesimals have grown in the inner
disk, since they would have a negligible opacity in the near-IR.  We note 
that \citet{cal05} reports the mass of the dust as 
7$\times$10$^{-10}$\,$\msun$; this is actually the mass of the gas within 
the hole, assuming the standard dust to gas mass ratio.  The gas mass could 
be significantly larger, depending on the total amount of solids and the 
actual ratio, but these are poorly constrained by existing data.

We vary the temperature of the wall to best reproduce the data.  The radius 
of the wall is set by the temperature and dust composition, and the wall's
height is set by the disk scale height.  We assume that the wall is 
axisymmetric and composed of relatively small grains, as well as vertically
flat in order to reproduce the rapid rise of the mid-IR excess at wavelengths 
beyond 10\,$\mu$m.  We adopt the dust composition used in \citet{dal05} and 
\citet{cal05}.  The maximum grain size is adjusted from ISM sizes to reproduce 
the shape of the IRS spectrum as necessary.  
At short wavelengths, larger grains have smaller opacities than 
ISM-sized grains. Therefore, at a given temperature large grains will
be at smaller radii than ISM-sized grains as per Eqn. 12 of \citet{dal05}.
The derived size of the inner hole varies somewhat depending on 
whether the SED or the resolved millimeter visibilities are included.  Fitting 
only the broadband SED and neglecting the resolved millimeter-wavelength data, 
the wall is located at 26\,AU and has a temperature of 130K and a height of 
$\sim$2\,AU with maximum grain size $a_{max}$=0.25\,{\micron} 
(Fig.~\ref{fig:catherine}, left panel).  The radius of the wall differs by 
$\sim$2 AU from \citet{cal05},  since here we take $L_{acc} \sim 
GM\mdot/R$ assuming magnetospheric accretion while \citet{cal05} uses 
$L_{acc} \sim GM\mdot/2R$ as per the boundary layer model.  We also
adopt a different mass accretion rate.

In order to compare the SED model with the resolved continuum data, it is
necessary to fix the disk geometry.  As listed in Table \ref{tab:prop}, we
adopt an inclination of 55$^\circ$, in order to maintain consistency with
\citet{cal05}.  However, the position angle is poorly reproduced by the 
value of $53.4^\circ \pm 0.9^\circ $ that is the weighted average of fits to 
the CO emission \citep[][see Fig.~\ref{fig:continuum}]{dut98,dut08}.  To 
derive a more appropriate position angle, we generate a sky-projected image 
from the disk model and use the MIRIAD task uvmodel to sample the image at 
the same spatial frequencies as the data.  We compare these model visibilities 
with the observed 860\,$\mu$m visibilities (which have the finest resolution).  
We repeat this process for a range of position angles and calculate a $\chi^2$ 
value comparing each set of model visibilities with the data.  Using this 
method, we fit a position angle of $64^\circ \pm 2^\circ$, which differs
by $11^\circ \pm 2^\circ$ from the position angle of the CO disk derived
by \citet{dut98,dut08}. 

When considering the resolved millimeter-wavelength visibilities, a disk 
with a 20\,AU hole reproduces the emission much better 
(Fig.~\ref{fig:catherine}, right panel, and Fig.~\ref{fig:continuum}, 
center panels).  Using the same $\chi^2$ comparison of visibilities as 
described in the previous paragraph, the 20\,AU model represents a 3$\sigma$
improvement over the 26\,AU model, which significantly underpredicts the
amount of flux produced close to the star.  This 20\,AU hole has a 
wall with a temperature of 120\,K, a height of 1.4\,AU, and maximum grain 
size $a_{max}$=5\,{\micron}.  For neither the 20\,AU nor the 26\,AU model 
does the wall contribute significant continuum emission at the wavelengths 
and spatial scales probed by our data.  The main discrepancy between the 
fits to the SED and the millimeter visibilities occurs between wavelengths 
of $\sim$20--40\,{\micron} where the 20\,AU hole model overpredicts the flux.  
However, the SED morphology in this region is likely sensitive to the 
properties of the wall at the inner disk edge, which are not well known and 
are not constrained by our data.  It is also possible that the composition of 
the grains, particularly whether the silicate and graphite form composite 
grains or are separated, can affect the temperature and therefore the 
mid-IR morphology of the wall component of the SED \citep{dal09}.  Since 
our focus is on the interferometric millimeter-wavelength data, we adopt 
the model with a 20\,AU inner hole for the remainder of the analysis.  
Figure~\ref{fig:continuum} compares this model with the data in the image 
plane (center panel) and in the visibility domain (red line in the right 
panel).  The agreement is excellent, and the residuals are less than 
3$\sigma$ within the 2\arcsec\ box shown.  

The flux density of the eastern peak of the 860\,$\mu$m image is 
50\,mJy\,beam$^{-1}$, while that of the western peak is 59\,mJy\,beam$^{-1}$.  
The corresponding peaks in the model images are 49 and 50\,mJy\,beam$^{-1}$,
respectively.  Given the rms noise of 3.5\,mJy\,beam$^{-1}$, these values are 
consistent with no flux difference and hence axially symmetric emission from 
the inner disk edge.  The positional accuracy of the data and knowledge of
the stellar proper motion are insufficient to determine whether or not the 
emission peaks are equally offset from the star.  This result may be 
contrasted with the strong asymmetries observed by \citet{bro08} in their 
observations of the inner hole in LkH$\alpha$ 330, although these data are
missing short antenna spacings present in the GM Aur data that may dilute 
asymmetries.  However, as in the case of LkH$\alpha$ 330, we find that the 
GM Aur continuum presents a sharp contrast in brightness between the inner 
and outer disk, reflected by the null in the visibility function and the 
strong agreement between the data and the model containing an inner hole.  
The $1.1 \times 10^{-11}$\,M$_\sun$ of dust within the central hole in the 
model implies a reduction in the mass surface density of small grains of at 
least 6 orders of magnitude at 1\,AU relative to a continuous model of the 
dust disk, indicating that the data are consistent with an inner disk region 
that is essentially completely evacuated of small grains.

\begin{table}
\caption{Stellar and Model Properties\label{tab:prop}}
\begin{tabular}{l c}
\hline
\multicolumn{2}{c}{Star$^{1}$}\\
\hline
$L_{*}$ $(L_{\sun})$.................. & 1.1  \\
$R_{*}$ $(R_{\sun})$.................. & 1.5  \\
$T_{*}$ $(K)$.................. & 4730 \\
$\mdot$ $(M_{\sun} yr^{-1})$.................. & 7.9 $\times$ 10$^{-9}$ \\
Distance (pc).................. & 140\\
$A_{V}$.................. & 1.2 \\
Inclination (deg).................. & 55\\
\hline
\multicolumn{2}{c}{Optically Thick Wall$^{2}$}\\
\hline
$R_{wall}$ $(AU)$.................. & 20 (26)  \\
$a_{min}$ ({\micron})$^{1}$.................. & 0.005  \\
$a_{max}$ ({\micron})$^{3}$.................. & 5 (0.25)  \\
$T_{wall}$ $(K)$$^{3}$.................. & 120 (130) \\
$z_{wall}$ $(AU)^{3,4}$.................. &  1.4 (2) \\
\hline
\multicolumn{2}{c}{Optically Thick Outer Disk}\\
\hline
$R_{d,out}$ $(AU)$$^{1}$.................. & 300 \\
$\epsilon$$^{3}$.................. & 0.5 \\
$\alpha$$^{3}$.................. & 0.002 \\
$M_{d}$ $(M_{\sun})$.................. & 0.16 \\
\hline
\end{tabular}
\tablenotetext{1}{These values are adopted.  Refer to text for references.}
\tablenotetext{2}{Values in parenthesis refer to parameters in the
case that the hole is 26 AU.}
\tablenotetext{3}{These are free parameters that are constrained by the
SED.}
\tablenotetext{4}{z$_{wall}$ is the height of the wall above the midplane}
\end{table}

\subsection{Comparison with CO Observations}
\label{sec:modelCO}

In order to compare the gas and dust properties of the GM Aur disk, we used
the SED-based model described above to generate predicted CO $J$=3$-$2 and 
$J$=2$-$1 emission.  We assume that gas and dust are well mixed, with a 
uniform gas-to-dust mass ratio of 100 (neglecting the complication of dust 
settling) and a constant CO abundance relative to H$_2$ of 10$^{-6}$, which 
is required to reproduce the peak CO $J$=2$-$1 flux.  We also add 
microturbulence with a FWHM of 0.17\,km\,s$^{-1}$ throughout the outer disk, 
as derived by \citet{dut98}.  This is comparable to the 0.18 km\,s$^{-1}$ 
spectral resolution of the data and does not affect our determination 
of the disk geometry.  Due to the position angle differences evident between 
the continuum emission in Fig.~\ref{fig:continuum} and the central channels in 
Fig.~\ref{fig:co32}, we also adjust the position angle to 51\degr\ \citep[as 
in ][]{dut98}.  Finally, we note that with an outer radius of 300\,AU, the 
continuum model severely underpredicts the CO emission at large radii, as 
expected for a model with a sharp cutoff at its outer edge \citep{hug08}.  
We therefore extrapolate the model to 525\,AU to match the spatial extent 
of the CO emission \citep{dut98}.  While this larger CO model no longer 
matches perfectly the continuum emission for the shortest baselines, based 
on the prediction assuming a constant gas-to-dust mass ratio, it retains the 
kinematic and thermal structure of the small-scale continuum model.  In 
order to consistently solve for the level populations and generate 
sky-projected images in the CO lines, we use the Monte Carlo radiative 
transfer code RATRAN \citep{hog00}.  We then use the MIRIAD task uvmodel 
to sample the model image at identical spatial frequencies to those present 
in our interferometric CO data set.  

\begin{figure*}[ht]
\epsscale{1.0}
\plotone{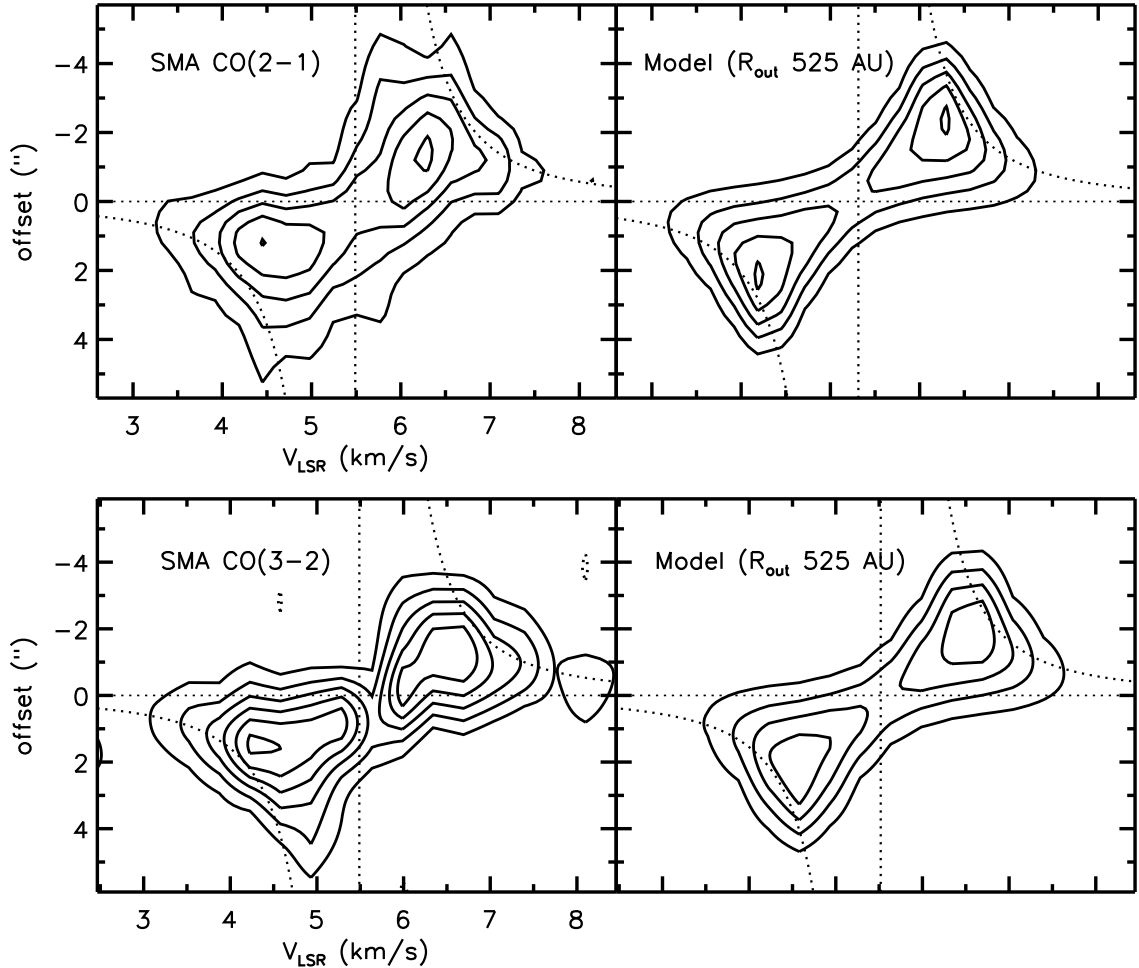}
\figcaption{
Position-velocity diagram comparing the molecular line observations 
({\it left}) with the predicted ({\it right}) CO $J$=2--1 ({\it top}) and 
CO $J$=3--2 ({\it bottom}) emission from the GM Aur disk, assuming a standard 
gas-to-dust mass ratio of 100.  The plots show the brightness as a function 
of distance along the disk major axis, assuming a position angle of 51\degr.  
Contours are [2,4,6,...] times the rms flux density in each map (0.17 and 
0.61 Jy beam$^{-1}$, respectively).  The dotted line shows the expected 
Keplerian rotation curve for a star of mass 0.84\,M$_\sun$.  The outer 
radius of the model has been extended to 525\,AU to reproduce the extent of 
the molecular gas emission (see \S\ref{sec:modelCO} for details).  The CO 
morphology is consistent with the SED-based model, with the exception of the 
line ratio: the model that best reproduces the peak flux of the CO $J$=2--1 
line underpredicts the CO $J$=3--2 brightness by 30\%. 
\label{fig:pv}}
\end{figure*}

Figure~\ref{fig:pv} compares the predicted CO emission from the extended 
SED model (right) with the observed emission from the GM Aur disk (left) for 
the $J$=2$-$1 (top) and $J$=3$-$2 transitions.  It is clear that the velocity 
pattern in the disk is consistent with Keplerian rotation \citep[as previously 
noted by ][]{koe93, dut98}, and that the SED-based model is capable of 
reproducing the basic morphology of the CO emission.  

The primary difference between data and model is the CO $J$=3--2/$J$=2--1 line 
ratio: the disk structure model that reproduces the peak flux density of 
the $J$=2$-$1 transition underpredicts the peak $J$=3$-$2 flux by 30\%.  
This difference may be attributed to a $\sim$10\,K difference in temperature 
between the gas and dust in the upper layers of the GM Aur disk that are 
probed by these optically thick CO lines.  While the vertical temperature 
gradient of the dust in the model is fixed by the SED, a relative increase 
in gas temperature would populate the upper rotational transition of the 
molecule more efficiently and produce more $J$=3$-$2 emission relative to 
$J$=2$-$1.  The temperature and the CO abundance are also somewhat 
interdependent, since the CO abundance sets the vertical location, and 
therefore the temperature, of the $\tau$=1 surface from which most of the 
line emission originates.  An increase in temperature would therefore 
also vary the anomalously low CO/H$_2$ ratio necessary to reproduce the 
$J$=2$-$1 flux.  Such line ratio differences have been previously observed 
in the disk around TW Hya \citep{qi04,qi06}, and may be due to additional 
heating of gas in the upper disk by such processes as x-ray and UV 
irradiation, dissociative or mechanical heating \citep[e.g.][]{gla04,kam04,
nom07}

Nevertheless, while the flux levels vary between the data and model 
prediction, the similarity in morphology makes it clear that the overall 
disk structure is consistent between the molecular gas traced by 
CO and the model based on dust traced by continuum emission and the SED.  
The only other significant difference between the two is in the position 
angle of the emission, which differs by $\sim$11\degr.  The implications of 
this result are discussed in \S\ref{sec:PA} below.

\section{Discussion}
\label{sec:discussion}

\subsection{Inner Disk Clearing}

The resolved millimeter continuum observations of the GM Aur system are 
consistent with the prediction from the SED model.  Models of the observed 
860\,$\mu$m and 1.3\,mm maps in conjunction with the SED and {\it Spitzer} 
IRS spectrum, give a value of $\sim$20 AU for the extent of this inner 
cleared region.  The inference of an inner 
hole of this size from the SED and resolved millimeter visibilities is 
consistent with recent millimeter-wave observations of rotational transitions 
of CO isotopologues from the GM Aur disk that provide spectroscopic evidence 
for a diminished density of cold CO within 20 AU \citep{dut08}.  However, 
other observations indicate that this region cannot be entirely devoid of 
gas.  \citet{sal07} detect CO rovibrational emission originating from 
hot gas at radii near $\sim0.5$\,AU, from which they infer a total gas mass 
in the inner disk of $\sim0.3 \textrm{~M}_{\earth}$.  Measurements of the 
H$\alpha$ linewidth imply an accretion rate of $\sim 10^{-8} 
\textrm{~M}_\sun$~yr$^{-1}$ \citep{whi01,ing09}; accretion at this rate 
requires a steady supply of gas from the inner disk.  The SED model also 
requires $3 \times 10^{-4}$ lunar masses of dust in the inner disk, to 
account for the 10\,$\mu$m silicate feature and slight near- to mid-IR 
excess \citep{cal05}.  

A wide variety of mechanisms has been invoked to explain the low optical 
depth of the central regions of transition disks \citep[see e.g.][and 
references therein]{naj07}, each with different implications for planet 
formation and the process of evolution between the primordial and debris 
disk stages.  The available measurements of properties of the inner hole 
in the GM Aur disk allow us to evaluate the plausibility of each mechanism 
as the driver of disk clearing in this system.  

{\it Grain Growth --} The agglomeration of dust into larger particles should 
proceed faster in central regions where relative velocities of particles are 
faster and surface densities are higher.  This would produce a drop in 
opacities associated only with the inefficiency of emission of large grains 
at the observed wavelengths \citep[e.g.][]{str89,dul05}.  However, this 
process is inconsistent with the clearing of CO from the central region 
observed by \citet{dut08}, as grain growth should proceed without diminishing 
the gas density.  Grain growth is also somewhat inconsistent with the steep 
submillimeter slope observed by \citet{rod06} for the GM Aur system.  The 
value inferred for the millimeter wavelength slope $\alpha$ of 3.2 is the 
steepest in their sample of ten T Tauri stars, and is typical of a grain
population that has undergone little growth, with grain size $a_{max} \le 
1$\,mm.  Furthermore, the original SED model and the submillimeter visibilities
both independently indicate a {\it sharp} decrease in surface density or 
opacity near 24\,AU, while grain growth and dust settling are predicted to 
be a continuous process and so should display a more gradual transition 
between the inner and outer disk \citep{wei97,dul05}.

{\it Photoevaporation --} Another proposed process to generate inside-out 
clearing of protoplanetary disks is photoevaporation via the ``UV switch'' 
mechanism \citep{cla01}.  In this scenario, high-energy photons from the 
star heat the upper disk layers, allowing material to escape the system 
at a rate that gradually diminishes the disk mass, while most of the disk 
mass drains onto the star via viscous accretion \citep[e.g.][]{har98}.  
Once the photoevaporation rate matches the accretion rate near 1~AU and 
prevents resupply of material from the outer disk, the inner disk will 
decouple and drain onto the star within a viscous timescale, leaving an 
evacuated central region surrounded by a low-mass outer disk that will 
then rapidly disperse.  As noted by \citet{ale07}, the properties of the 
GM Aur system are inconsistent with a photoevaporative scenario because the 
large mass of the outer disk should still be sufficient to provide a 
substantial accretion rate to counteract the photoevaporative wind.  
Furthermore, the measured accretion rate is high enough that within the 
framework of the photoevaporation scenario, it would only be observed during 
the brief period of time when the inner disk was draining onto the star.  
Photoevaporation may yet play a role in clearing the outer disk of its 
remaining gas and dust, but it cannot explain the current lack of inner 
disk material.

{\it Inside-Out MRI Clearing --} The magnetorotational instability operating 
on the inner disk edge may also drive accretion and central clearing, although
it should be noted that this is purely an evacuation mechanism: it can only 
take hold after the generation of a gap by some other means. 
Nevertheless, given the creation of a gap, MRI clearing is 
predicted to operate in systems like GM Aur whose outer disks are still too 
massive for photoevaporation to dominate \citep{chi07}.  
The observed depletion of CO interior to 20 AU radius \citep{dut08} is 
consistent with this theory, which predicts a total gas mass depletion 
of order 1000$\times$ interior to the rim radius relative to the extrapolated 
value from the outer disk power law fit, normalizing to the total disk mass of 
0.16$M_\sun$.  This theory is consistent with the substantial accretion rate 
of the GM Aur system, yielding a value of $\alpha$ of 0.005, only slightly 
greater than the derived value of 0.002 from the model.  \citet{sal07} 
estimate a gas-to-dust ratio of $\sim1000$ in the inner disk, roughly 10 
times greater than that of the outer disk, which is consistent with the 
prediction of the inside-out MRI evaporation scenario that flux from the 
star should promote blowout of small dust grains by radiation pressure, 
substantially clearing the inner disk of dust even as the gas continues 
to accrete onto the star.  However, it is difficult to reconcile this 
with the substantial population of $\mu$m-size grains that must be present 
in the inner disk to account for the 10\,$\mu$m silicate feature in the 
IRS spectrum.  It is also important to consider the source of the requisite
initial gap in the disk.

{\it Binarity --} The dynamical influence of an unseen stellar or substellar 
companion would also cause clearing of the inner disk.  A notable 
example is the recent result by \citet{ire08} demonstrating that the inner 
hole in the transition disk around CoKu Tau/4 is caused by a previously 
unobserved companion.  There are relatively few constraints on the 
multiplicity of GM Aur at the $<20$\,AU separations relevant for the 
inner hole.  Radial velocity studies with km\,s$^{-1}$ precision do not 
note variability \citep{bou86,har86}, ruling out a close massive companion.  
As \citet{dut08} discuss, the stellar temperature and dynamical mass from the
disk rotation combined with the H-band flux place an upper limit of 
$\sim$0.3 M$_\sun$ on the mass of a companion.  Interferometric 
aperture-masking observations with NIRC2 that take advantage of adaptive 
optics on the Keck II telescope place an upper limit of $\sim$40 
times the mass of Jupiter on companions with separations between 1.5 and 
35\,AU from the primary (A. Kraus and M. Ireland, private communication).  
The presence of hot CO in the central 1\,AU of the system \citep{sal07} and
the high accretion rate, undiminished relative to the Taurus median, also 
argue against the presence of a massive close companion.  A stellar 
companion is therefore an unlikely origin for the central clearing in the 
GM Aur system.

{\it Planet-Disk Interaction --} Perhaps the most compelling mechanism for 
producing a transition disk is the dynamic clearing of material by a giant 
planet a few times the mass of Jupiter.  The opening of gaps and 
holes in circumstellar disks has long been predicted as a consequence of giant 
planet formation \citep[e.g.][]{lin86,bry99}.  Some simulations have 
shown that inner holes may in fact be a more common outcome than gaps as 
angular momentum transfer mediated by spiral density waves can clear the 
inner disk faster than the viscous timescale \citep{var06,lub06}.  
The planet-induced clearing scenario was considered in detail for GM Aur 
by \citet{ric03} and found to be globally consistent with the observed 
properties of the system (although their estimate of the inner hole radius is 
based on pre-{\it Spitzer} SED information).  This mechanism naturally explains
the diminished but persistent accretion rates and presence of small dust grains
through two predictions of models of planet-disk interaction: (1) filtration 
of dust grains according to size is expected at the inner disk edge, leading 
to a dominant population of small grains in the inner disk \citep{ric06}; and
(2) a sustained reduction in accretion rate to $\sim10\%$ of that through the 
outer disk is predicted as the giant planet begins to intercept most of the 
accreting material \citep{lub06}.  These effects combined may also explain 
the enhanced gas-to-dust ratio in the inner disk.  A planet-induced gap 
could also serve as a catalyst for inside-out MRI clearing \citet{chi07}.

Given the observed 20\,AU inner disk radius and the scenario of clearing via
dynamical interaction with a giant planet, it is possible to make a simple
estimate of the distance of the planet from the star.  The width of a gap 
opened by a planet is approximately $2\sqrt{3}$ Roche radii \citep{art87},
and simulations show that the minimum mass necessary to open a gap is of
order 1 Jupiter mass \citep[e.g.][]{lin93,edg07}.  If the outer edge of the 
planet-induced gap coincides with the 20\,AU inner disk radius (with the 
portion of the disk interior to the planet cleared via spiral density waves 
or the MRI), then a companion between 1 and 40 times the mass of Jupiter 
would be located between 11 and 16\,AU from the star.  The influence of a 
planet carving out an inner cavity in the dust distribution is therefore a 
plausible scenario, bolstered by recent results demonstrating that a planet 
is responsible for dynamical sculpting of dust in the much older Fomalhaut 
system \citep{kal08}.

\subsection{Evidence for a Warp?}
\label{sec:PA}

While the model comparison in \S\ref{sec:model} above shows that CO emission 
from the disk is globally consistent with Keplerian rotation, the 11\degr\ 
difference in position angle between the continuum data and the two 
CO data sets is significant at the $\sim5\sigma$ level, and may indicate some 
kinematic deviation from pure Keplerian rotation in a single plane.  Changes 
in position angle with physical scale are commonly interpreted as warps in the 
context of studies of galaxy dynamics \citep[e.g.][]{rog74}; it may be that 
the change in position angle in the GM Aur disk indicates a kinematic warp.  

The possibility of a warp or other deviation from Keplerian rotation was
discussed by \citet{dut98}, although their discussion was based on possible
isophote twisting observed in integrated CO $J$=2$-$1 contours.  We observe no 
such isophote twisting in the integrated CO $J$=2--1 or $J$=3--2 emission 
presented here (Fig. \ref{fig:momfig}), although this determination may be 
influenced by the differing baseline lengths and beam shapes in the respective 
interferometric data sets.  Instead, we observe deviations from the expected 
position angle only in the rotation pattern of the resolved CO emission, 
which is reflected in the isovelocity contours of Fig. \ref{fig:momfig}.  
This position angle change does not appear to be related to the cloud 
contamination, as it is more clear in the less-contaminated CO $J$=3$-$2 
data set.  In order to test whether the position angle of the true brightness 
distribution might have been altered by incomplete sampling of the data in 
the Fourier domain, we generated a model of the disk at a position angle of 
64\degr, consistent with that measured independently for the two continuum 
data sets.  We then fit the position angle by $\chi^2$ minimization as in 
\S\ref{sec:modelcont} above.  With this method, after sampling with the response
at the spatial frequencies in the CO $J$=3--2 data set, we recover the position
angle to within less than a degree of the input model.  This is to be 
expected, since the $\chi^2$ fitting procedure takes into account the 
interferometer response when fitting for the position angle.  The position 
angle change is therefore robust independent of beam convolution effects.  

In order to cause a change in position angle on physical scales between those
probed by the continuum ($\sim 30$\,AU) and the CO ($\sim 200$\,AU), a warp
would have to occur at a size scale of order 100\,AU.  The most natural
explanations for the presence of a warp in a gas-rich circumstellar disk
include flybys and perturbations by a planet or substellar companion.  
A simple estimate of the timescale of flyby interactions is $\tau = 1 / (N 
\pi b^2 \sigma)$, where $N$ is the number density of stars, $b$ is the 
approach distance, and $\sigma$ is the velocity dispersion.  Assuming typical 
values for Taurus, including a stellar density of $\sim$10\,pc$^{-3}$ 
\citep[e.g.][]{gom93} and velocity dispersion of 0.2\,km\,s$^{-1}$ 
\citep{kra08}, the timescale for interactions at distances of $\sim$1000 AU, 
sufficient to cause significant perturbations at Oort Cloud radii 
\citep{sch82}, is of order 1\,Gyr.  Since the results of a one-time 
perturbation would likely damp in a few orbital periods (10$^3$\,yr at 
a distance of 100\,AU), such an interaction is statistically unlikely. 
However, it should be noted that a recent interaction might have been capable
of producing an extended feature like the ``blue ribbon'' observed in 
scattered light by \citet{sch03}.  

The influence of a massive planet or substellar companion has been investigated
as the origin of warps observed in gas-depleted debris disks, including 
$\beta$ Pic \citep{mou97} and HD~100546 \citep{qui06}.  However, there is
a dearth of theoretical investigation into the plausibility of warps caused
by planetary systems in gas-rich disks more closely analogous to the GM Aur
system.  Since the warp in the GM Aur disk must occur between the Hill sphere 
of the putative planet and the $\sim$200\,AU resolution of the CO line 
observations, it is plausible that the warp could be due to the 
gravitational influence of the same body responsible for evacuating the inner 
disk.  A theoretical inquiry into this possibility would be useful, but is
beyond the scope of this paper.

\section{Conclusions}

Spatially resolved observations in millimeter continuum emission, obtained
using the SMA at 860\,$\mu$m and PdBI at 1.3\,mm, reveal a sharp decrease 
in optical depth near the center of the GM Aur disk.  Simple estimates of the 
extent of this region, based on the separation of peaks in the continuum 
images and the position of the null in the visibility functions in 
Fig.~\ref{fig:continuum}, are consistent with the inner hole radius of 24\,AU
derived by \citet{cal05} using disk structure models to fit the SED.  No 
significant azimuthal asymmetry is detected in the continuum emission.

Refined versions of the SED-based model of \citet{cal05} show that the
data are very well reproduced by a disk model with an inner hole of radius 
20\,AU.  This model overpredicts the broadband SED flux in the 20--40\,$\mu$m
wavelength regime, but this region of the spectrum likely depends on the 
properties of the wall at the inner disk edge, which are poorly constrained
by available data.

CO emission in the $J$=3$-$2 and $J$=2$-$1 transitions confirms the presence
of a disk with kinematics consistent with Keplerian rotation about the 
central star, but at a position angle offset from the continuum by 
$\sim$11\degr.  The morphology of the CO emission is broadly consistent with 
the SED model, but with a larger CO $J$=3--2/$J$=2--1 line ratio than 
predicted for the SED model.  This is a likely indication of additional gas 
heating relative to dust in the upper disk atmosphere.

Given the observed properties of the GM Aur system, photoevaporation,
grain growth, and binarity are unlikely physical mechanisms for inducing 
a sharp decrease in opacity or surface density at the disk center.  The 
inner hole plausibly results from the dynamical influence of a planet on 
the disk material, with the inner disk possibly cleared by spiral density 
waves or the MRI.  While a recent flyby is statistically unlikely, warping 
induced by a planet could also explain the difference in position angle 
between the continuum and CO data sets.  

\acknowledgments{The authors would like to thank the IRAM staff, particularly
Roberto Neri, for their help with the observations and data reduction.  
We thank Lee Hartmann for helpful discussions in the early stages of this
project.  Partial support for this work was provided by NASA Origins of Solar 
Systems Program Grant NAG5-11777.  A.~M.~H. acknowledges support from a 
National Science Foundation Graduate Research Fellowship.  Support for 
S.~M.~A. was provided by NASA through Hubble Fellowship grant \#HF-01203-A 
awarded by the Space Telescope Science Institute, which is operated by the 
Association of Universities for Research in Astronomy, Inc., for NASA, under 
contract NAS 5-26555.  N.~C. acknowledges support from NASA Origins Grant 
NNG05GI26G and JPL grant AR50406.  P.~D. acknowledges grants from CONACyT, 
M{\'{e}}xico.  J.~P.~W. acknowledges support from NSF grant AST-0808144.
}

\end{document}